\documentclass[sigconf]{acmart}

\usepackage[utf8]{inputenc}
\usepackage{graphicx}
\usepackage{url}
\usepackage{natbib}
\usepackage{tikz}
\usepackage{booktabs}
\usepackage{color}
\usepackage{gensymb}

\AtBeginDocument{%
  \providecommand\BibTeX{{%
    \normalfont B\kern-0.5em{\scshape i\kern-0.25em b}\kern-0.8em\TeX}}}

\setcopyright{acmcopyright}
\copyrightyear{2022}
\acmYear{2022}

\copyrightyear{2022} 
\acmYear{2022} 
\setcopyright{rightsretained} 
\acmConference[WebSci '22]{14th ACM Web Science Conference 2022}{June 26--29, 2022}{Barcelona, Spain}
\acmBooktitle{14th ACM Web Science Conference 2022 (WebSci '22), June 26--29, 2022, Barcelona, Spain}
\acmDOI{10.1145/3501247.3531589}
\acmISBN{978-1-4503-9191-7/22/06}

\begin{document}

\title{Desaparecidxs: characterizing the population of missing children using Twitter}


 \author{Carolina Coimbra Vieira}
 \affiliation{%
   \institution{Max Planck Institute for Demographic Research}
   \city{Rostock}
   \country{Gemany}
 }
 \email{coimbravieira@demogr.mpg.de}
 
 \author{Diego Alburez-Gutierrez}
 \affiliation{%
   \institution{Max Planck Institute for Demographic Research}
   \city{Rostock}
   \country{Gemany}
 }
 \email{alburezgutierrez@demogr.mpg.de}
 
 \author{Mar\'{i}lia R. Nepomuceno}
 \affiliation{%
   \institution{Max Planck Institute for Demographic Research}
   \city{Rostock}
   \country{Gemany}
 }
 \email{nepomuceno@demogr.mpg.de}

 \author{Tom Theile}
 \affiliation{%
   \institution{Max Planck Institute for Demographic Research}
   \city{Rostock}
   \country{Gemany}
 }
 \email{theile@demogr.mpg.de}

\renewcommand{\shortauthors}{Vieira, et al.}

\begin{abstract}
Missing children, i.e., children reported to a relevant authority as having ``disappeared,'' constitute an important but often overlooked population. From a research perspective, missing children constitute a hard-to-reach population about which little is known. This is a particular problem in regions of the Global South that lack robust or centralized data collection systems. In this study, we analyze the composition of the population of missing children in Guatemala, a country with high levels of violence. We contrast the official aggregated-level data from the Guatemalan National Police during the 2018-2020 period with real-time individual-level data on missing children from the official Twitter account of the \textit{Alerta Alba-Keneth}, a governmental warning system tasked with disseminating information about missing children. Using the Twitter data, we characterize the population of missing children in Guatemala by single-year age, sex, and place of disappearance. Our results show that women are more likely to be reported as missing, particularly those aged 13-17. We discuss the findings in light of the known links between missing people, violence, and human trafficking. Finally, the study highlights the potential of web data to contribute to society by improving our understanding of this and similar hard-to-reach populations.
\end{abstract}




\maketitle

\section{Introduction}
\label{sec:intro}

Quantitative researchers often encounter `missing' values in their data, for instance, derived from measurement error or non-response.
`Missingness' in real life, however, is an altogether different phenomenon. 
Missing people are individuals, whose status as dead or alive is unknown, reported as missing by relatives and friends. 
The Wall Street Journal reported in 2012 that: ``It is estimated that some 8 million children go missing around the world each year''\footnote{\url{https://www.wsj.com/articles/SB10001424052702304707604577424451609727644}}. 
However, missing people have received surprisingly little attention in the literature. 
This may be partly due to the fact that governments do not always collect and release up-to-date data on missing people \cite{citroni_pitfalls_2014} and disappearances are relatively rare in high-income countries\footnote{\url{http://amberalert.eu/statistics/}}. 

Missing people are a particular concern in enclaves of the Global South in which populations are exposed to high levels of economic uncertainty, poverty, violence, and conflict \cite{mcilwaine2001violence,soares2011homicide,meneghel2011femicides,garcia2019impact}.
We focus on the case of Guatemala, which emerged from a bloody 36-year civil war in 1996 \cite{ceh_capitulo_1999}.
Over the last decades, the country has experienced growing levels of violence related to gang activity and drug trafficking \cite{cruz2020gang} that alongside high levels of poverty, have contributed to large-scale population displacement and migration \cite{jonas2015guatemala}. 

Aggregated records from the National Civilian Police are one of the few data on missing people in Guatemala. 
According to these data, almost 40,000 individuals went missing in the country between 2003 and 2020, half of whom were under 18 years of age.
Between 2018 and 2020, the period we focus on in this study, 4,000 individuals went missing in Guatemala, 65\% of whom were children (0-17 years old). 
The police data, however, provide neither a full nor an updated picture of the population of missing people. 
The reasons why a person disappears may also be associated with age, gender, race, socioeconomic level, and contextual factors. However, little is still known about the characteristics of the missing people.

In this work, we adopt a methodology for collecting Twitter data in a systematic way, including image processing techniques to extract text from images, to track the population of missing children in Guatemala in real-time. We collected individual-level data of more than seven hundred missing children from the official Twitter account of the \emph{Alerta Alba-Keneth}, a governmental warning system tasked with disseminating information about missing children in Guatemala \cite{rodas_andrade_informe_2021}.


We combine the Twitter data with official data sources to provide the first systematic description of the population of missing children in Guatemala during the 2018-2020 period. This work addresses the following research question: \textbf{What is the composition of the population of missing children in Guatemala by age, sex, and geography, and how has this varied over time?}

To the best of our knowledge, this is the first work studying the demographic composition of missing children using Twitter data. We focus on Guatemala, but the methodology we adopt in this work can be replicated for other Twitter accounts sharing detailed information about missing people. Moreover, this work highlights the advantages of combining  social media data combined with official government data to study phenomena of societal relevance.

The rest of the paper is organized as follows. In Section \ref{sec:related}, we describe some related work. In Section~\ref{sec:data}, we detail the methodology used to collect and pre-process the data. The results are presented in Section \ref{sec:results}. In Section~\ref{sec:discussion}, we present the discussion of our results and offer additional comments about the applicability, limitations, and future work on using social media data to study missing children. Finally, our conclusions are presented in Section \ref{sec:conclusion}.

\section{Related Work}
\label{sec:related}


Existing studies have documented how individuals affected by the disappearance of an acquaintance use social networks to crowdsource support. 
Despite the fact that there is no guarantee that reporting missing people online will lead to their discovery, social networks provide much needed emotional support to friends and relatives of the missing people \cite{hattingh_using_2016}. Police departments around the world increasingly use social media to engage the surrounding communities using social media platforms \cite{dai2017working}.

Exploratory studies have examined the use of social media tools by police departments, with a particular focus on Twitter ~\cite{crump_what_2011,ferguson_missing_2021,solymosi_exploring_2021}. 
Most of these studies have centered on understanding how to increase the public's engagement with the information on missing people shared online. 
In \cite{ferguson_missing_2021}, the authors analyzed 373 missing person tweets posted over two years (2017–2019) from 15 Canadian police services on Twitter to estimate which features are likely to increase public engagement (retweets, likes, and comments) with these tweets. Similarly, \cite{solymosi_exploring_2021} analyzed 1008 Tweets made by Greater Manchester Police between the period of 2011 and 2018 in order to investigate what features of the tweet, the Twitter account, and the missing person are associated with levels of retweeting. In both studies, the authors found several features to be significantly associated with higher engagement, such as the use of images and hashtags.
These strategies increased community outreach and participation, as well as the likelihood of efficiently and successfully solving the missing person cases. 
A standardized structure for sharing details of missing people on Twitter may enhance the usefulness of social media in this respect \cite{ferguson_missing_2021}.

In this work, we collect information about missing children from the official Twitter account of \textit{Alerta Alba-Keneth}, a governmental warning system focused on missing children in Guatemala. The main goal of this work is to characterize the population of missing children, instead of predicting the best features responsible for a high engagement with tweets. Moreover, we do not expect variations between engagement due to different features used to tweet (e.g., the use of different hashtags) since the \textit{Alerta Alba-Keneth} Twitter account is highly structured and all the tweets and images shared have standardized components.

Finally, by focusing on the characterization of the missing children in Guatemala, we expect to provide a demographic overview of the missing child, minimizing the race- and gender-related media bias (e.g., African American missing children and female missing children being significantly underrepresented in television news coverage) as reported by other authors \cite{jae_missing_2010}.
To the best of our knowledge, this is the first work studying the demographic composition of missing children using Twitter data. 

\section{Data}
\label{sec:data}

\subsection{Guatemalan National Police data}
Missing people in Guatemala can be directly reported to three government offices: the National Civilian Police (Policía Nacional Civil), the Prosecutor General's Office (Ministerio Público), or the Attorney General's Office (Procuraduría General de la Nación). Each of these offices produces its own statistics, which means that reports of disappearances are often duplicated across sources. In our experience, accessing these data is very difficult given the unwillingness of the relevant authorities to share it. 

We obtained data on missing people in Guatemala via multiple Freedom of Information Requests to the National Civilian Police. We made similar requests to a number of other government agencies that collect reports of missing people, but this information was refused. The data from the National Civilian Police cover the 2003-2021 period, but our analyses are limited to the 2018-2020 period. The National Police data only includes disappearances reported to this institution directly.

According to the National Police, the population under 18 years of age represents more than half of the known cases of missing people in the country. These data include the number of missing people reported to the police by broad age groups (0-17; 18-35; 36-55; 56+) and month of occurrence. However, the police authorities did not share individual-level data or aggregated information on the composition of the population by age, sex, or ethnicity. Other details about the circumstances of the events (e.g., place of disappearance) were also unavailable. 

\subsection{Twitter data}
Twitter is a popular microblogging platform that allows users to express themselves and record their thoughts in 280 characters at maximum. Twitter is also a social network since it is possible to follow users to be updated about their tweets (which may contain text, photos, GIFs, videos, and links). Among its more than 200 million daily active users\footnote{\url{https://s22.q4cdn.com/826641620/files/doc_financials/2021/q2/Q2'21_InvestorFactSheet.pdf}} are news outlets, academic institutions, and government agencies who use the platform to share official information with a wide audience. We are particularly interested in the latter’s use of Twitter to share information on missing people. The data provided by Twitter has been used extensively by researchers in social science \cite{mccormick2017using,tsoi2018twitterdementia}.

We are interested in leveraging the use of Twitter by a Guatemalan government agency to gather detailed data on missing children, i.e. those under 18 years old according to Guatemalan law. We focus on the \textit{Alerta Alba-Keneth}\footnote{Official website available here: \url{https://www.albakeneth.gob.gt/}}, an inter-governmental agency tasked with the search, location, and immediate protection of disappeared or abducted children and adolescents. \textit{Alerta Alba-Keneth} was established in 2010 to coordinate the efforts of multiple government agencies for locating missing children, including the National Civilian Police, the Prosecutor General's Office, and the Attorney General's Office. It collects reports on missing children made to any of these agencies in addition to those made directly to \textit{Alerta Alba-Keneth} through their website and hotline, making it the most comprehensive source of data on missing children in the country \cite{rodas_andrade_informe_2021}.

\begin{figure}[!ht]
     \centering
     \includegraphics[trim = 0cm 0.1cm 0.1cm 0cm, clip=True, width=0.65\linewidth]{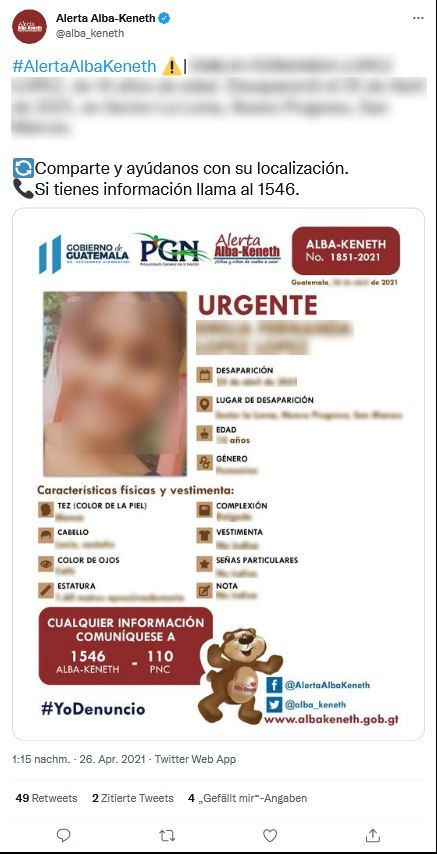}
     \caption{Example of a tweet from the \textit{Alerta Alba-Keneth} Twitter account (@alba\_keneth).}
     \label{fig:tweet}
\end{figure}

The \textit{Alerta Alba-Keneth} Twitter account, set up in 2015, has high visibility with almost 20 thousand followers. The \textit{@alba\_keneth} profile tweets images with information about newly missing children on a daily basis, following the structure of the image shown in Figure \ref{fig:tweet}.

We used the Twitter API Academic Research product track\footnote{\url{https://developer.twitter.com/en/products/twitter-api/academic-research}} to download all the tweets and their embedded images of the \textit{Alerta Alba-Keneth} account since 2015. Regular tweets about missing children in a consistent format started in 2018. Every tweet consists of a short text and an image. The text contains information about the name, age, date, and place of disappearance. The image contains not only a portrait photo of the missing child, but also a "profile" with more information in text form, including sex, appearance or individual characteristics (e.g., eyes color, hair color, and height) and contact information.

We collected a total of 13,696 tweets containing images. Considering the fact that some images are tweeted more than one time, by comparing the \textit{hash}\footnote{\url{https://docs.python.org/3/library/hashlib.html}} of each image, all duplicated images have been removed. We processed all these 11,130 unique images to extract information related to the demographic attributes of the sample. We used Optical Character Recognition (OCR) via the Python package PyTesseract to extract information on the children's demographic attributes (date and place of disappearance, age, and sex) from images. This information was not always available in the tweet text and when it was included, it was in an unstructured form. Thus it was easier to extract this information from the structured images. At the end of this process, we collected structured information about 7,800 unique missing children. Of 7,800 unique missing children, 6,875 correspond to tweets from 2018 or later and as such included in our study.

In order to extract the coordinates of the place of disappearance, we used the address reported in the ``place of disappearance'' field as input for the two geocoding services of Bing\footnote{\url{https://docs.microsoft.com/en-us/bingmaps/rest-services/}} and ArcGIS\footnote{\url{https://geocode.arcgis.com/arcgis/}}. We calculated the Haversine distance between the two results for an extra description of the accuracy of the address geolocation.

\subsection{Data protection}
All data used in this study have been made publicly available on Twitter by the \textit{Alerta Alba-Keneth} account. Nevertheless, data protection was a major concern when collecting and processing Twitter data. We recognize that the data are sensitive because they concern minors going through a difficult situation that also affects their families. For this reason, we kept the data in an encrypted drive and, even though we conducted individual-level analysis, this article includes only aggregated results. Moreover, in the text, we eliminated all personally-identifying information so that no child or relative could be uniquely identified.

\section{Results}
\label{sec:results}

In this section, we explore time and demographic trends of missing children reports in Guatemala for the 2018-2020 period. The main results come from Twitter data from the \textit{Alerta Alba-Keneth} account. Whenever possible, we compare these results to ``official'' data obtained from the National Police. It is worth emphasizing that although both sources refer to the same phenomenon -- missing children reports -- we do not expect them to match perfectly. This is due to the fact that both institutions have independent, though complementary, data collection systems, as explained above. Nevertheless, we expect that both data sources will reflect similar trends over time, even if absolute numbers differ.

\begin{figure}[!ht]
    \centering
    \includegraphics[width=0.7\linewidth]{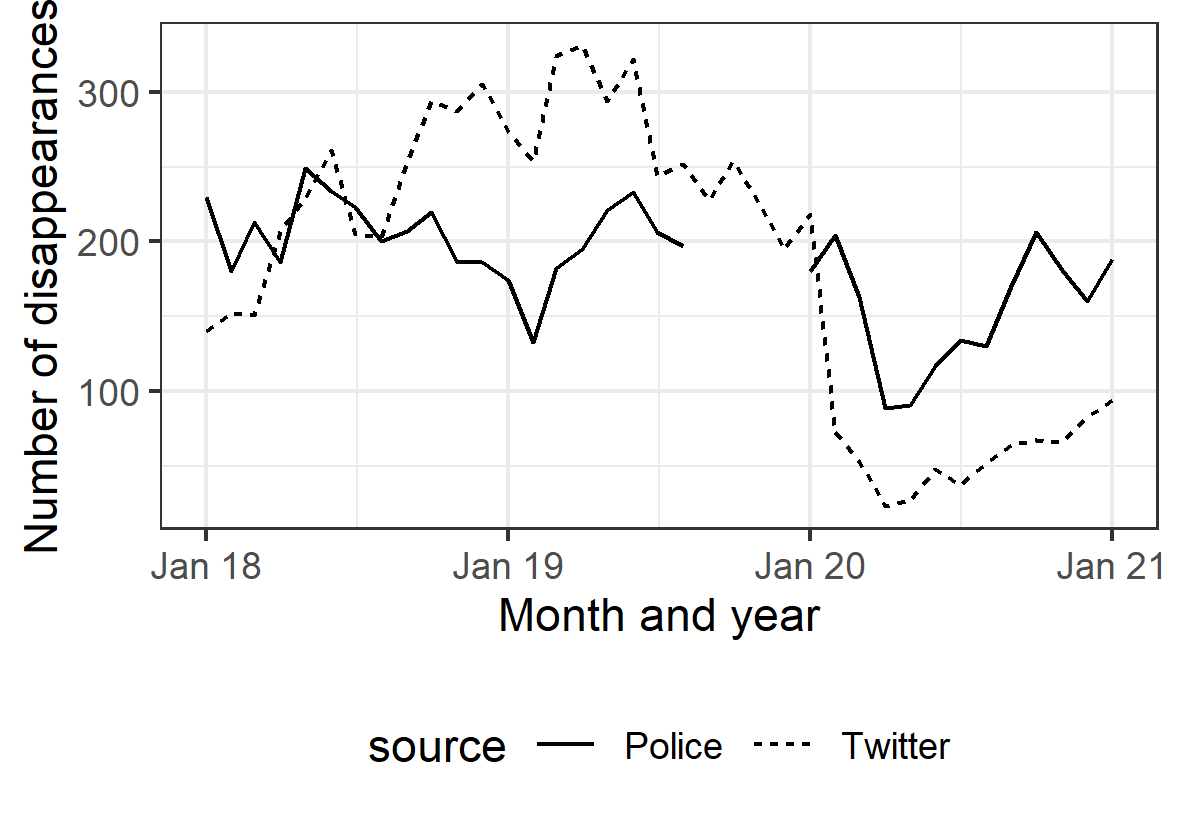}
    \caption{Number of missing children by month: a comparative overview of data from the Guatemalan National Police and \textit{Alerta Alba-Keneth} Twitter data. Note that data for May-December 2019 were not provided by the National Police.}
    \label{fig:police-twitter}
\end{figure}

As a first step, we contrast the number of missing children reported by the Twitter data to the official police data over the period of study. Figure \ref{fig:police-twitter} shows the number of children reported as missing according to both data sources. The image shows a general correspondence in the general trends of child disappearances over time. Although both series show similar fluctuations over time, the number of disappearances is higher in the Twitter data for most of 2019 and lower for 2020. Note that the latter does not necessarily reflect a falling number of disappearances -- the decline may be related to the Covid-19 pandemic \citep{martinez-folgar_excess_2021}. According to a recent report by the Ombudsman of Guatemala \citep{rodas_andrade_informe_2021}, strict lockdowns, more restricted opening hours, and reduced access to public transportation may all have contributed to limiting the number of missing children reports received by \textit{Alerta Alba-Keneth} in 2020. 

The measure presented in the paragraph above can be regarded as a rough measure of incidence. That is, the monthly number of disappearances is a proxy of how common it is for children to go missing. An advantage of the individual-level Twitter data is that it allows us to desegregate the population of missing children by age and sex. The police data do not include information on the distribution of missing children for different combinations of age and sex. This is crucial in order to identify the populations at the highest risk of disappearances and is particularly useful given the known links between missing children and human trafficking \citep{rodas_andrade_informe_2021}.

\begin{figure}[!ht]
    \centering
    \includegraphics[width=0.75\linewidth]{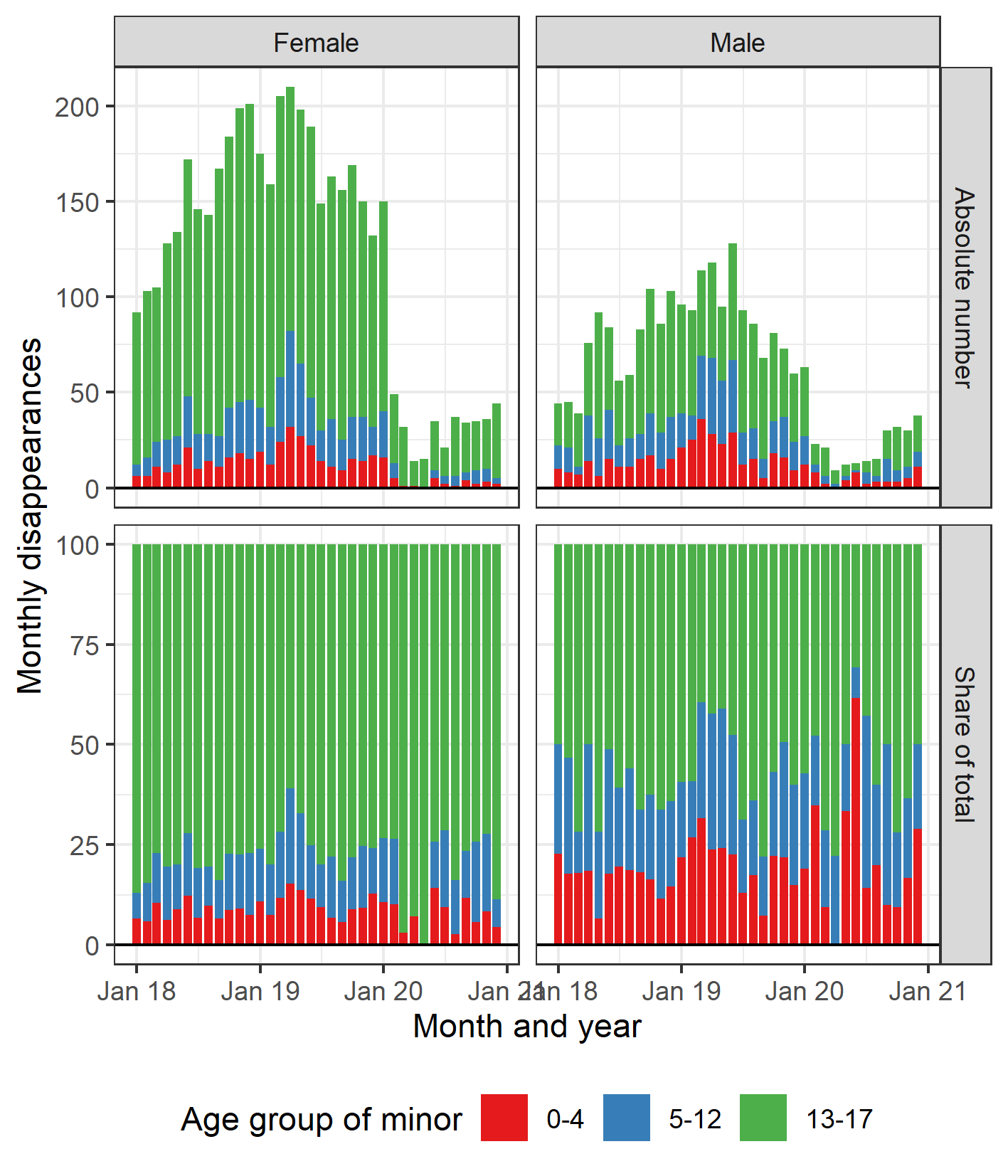}
    \caption{Age and sex distribution of the number of  missing children by month of reported disappearance (2018 - 2020) according to the \textit{Alerta Alba-Keneth} Twitter data.}
    \label{fig:twitter-age}
\end{figure}

Figure \ref{fig:twitter-age} shows the distribution of the missing minors by three age sub-groups intended to represent young children (0-4), children (4-12), and adolescents (13-17). 
The top panels show the absolute number of monthly reports of missing children and the bottom panels show the age composition of the population of missing children every month.
The first thing to note is that the age distributions remain relatively constant over time (bottom panels) even as the absolute number of reported cases fluctuates visibly over time. 
Indeed, this is true even for the 2020 period, in which the number of missing children reports was heavily affected by the Covid-19 pandemic \citep{rodas_andrade_informe_2021}.
The figure also shows that adolescents constitute the largest group of missing children in the period we study.
The proportions of young children and children are both considerably lower and similar to each other.

The most striking finding of Figure \ref{fig:twitter-age} is the disparities by sex. 
Focusing on the bottom panels, we see that there are important differences between the age distribution of missing male and female children. 
Concretely, missing girls tend to be considerably older than missing boys--around 75\% of missing girls were over 13 years old whereas only about 60\% of missing boys were adolescents.
Female adolescents are the largest sub-population of missing children and were almost twice as likely to be reported as missing as boys the same age.

\begin{figure}[!ht]
 \centering
 \includegraphics[width=0.8\linewidth]{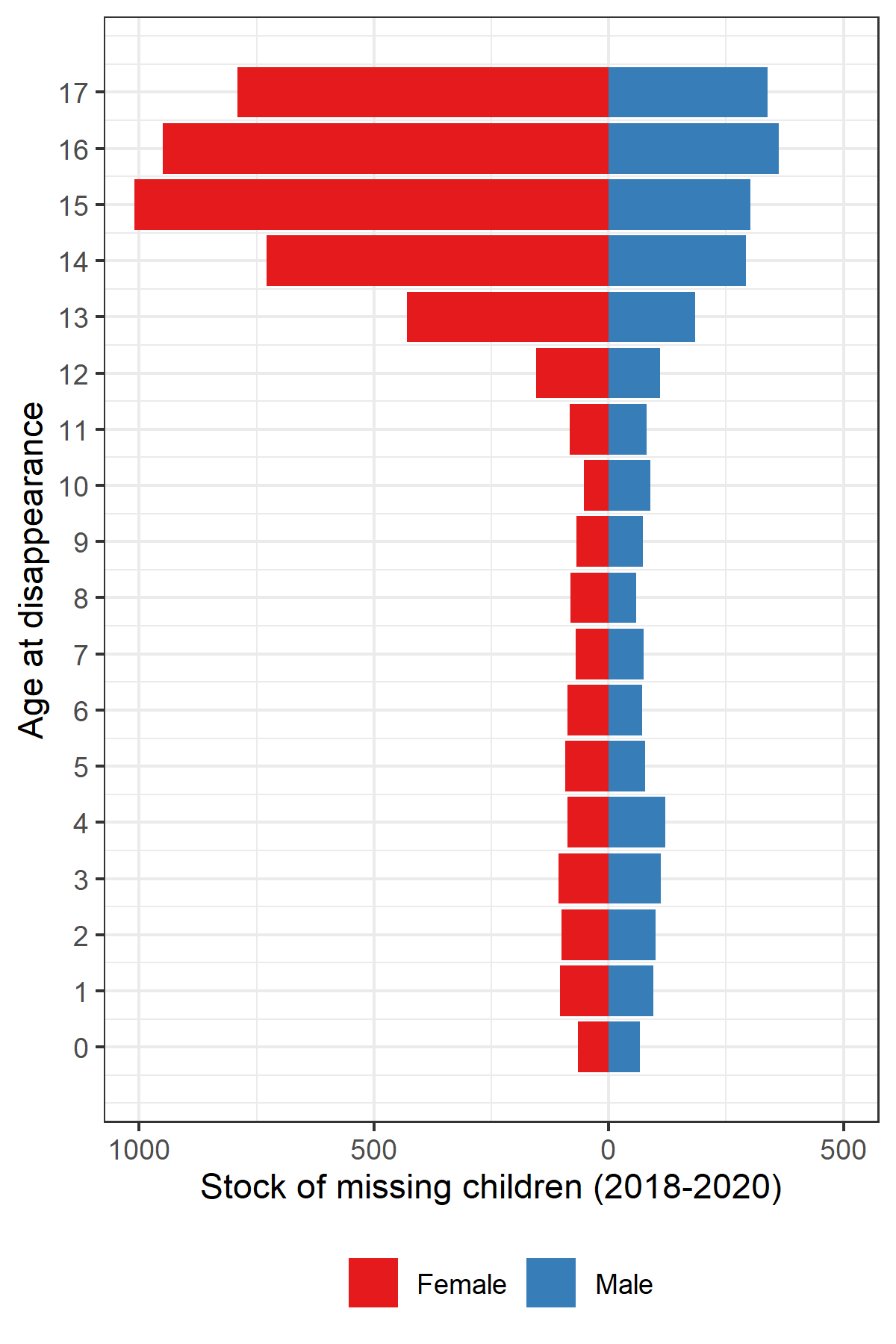}
 \caption{Stock of missing children: cumulative number of disappearances (2018 - 2020) according to the \textit{Alerta Alba-Keneth} Twitter data.}
    \label{fig:twitter-sex}
\end{figure}

Figure \ref{fig:twitter-sex} summarizes the stock of missing children in the 2018-2020 period by age and sex. The image confirms the patterns discussed above and shows the large disparities in the number of missing male and female children. The number of women aged between 10 and 17 years was 4,200, more than twice as high as the number of men reported as missing in the same age group (1,800). As expected, the ratio between the number of adolescent women and men is largest for older ages. There are 2.7 times more women aged 15-17 reported as missing than men of the same age. This large gap between sexes is absent for younger age groups (e.g., 870 girls younger than 10 years were reported as missing, compared to 850 boys).

\begin{figure*}[!ht]
    \centering
    \includegraphics[trim=1cm 1.5cm 0.7cm 0cm, clip=true,width=0.55\linewidth]{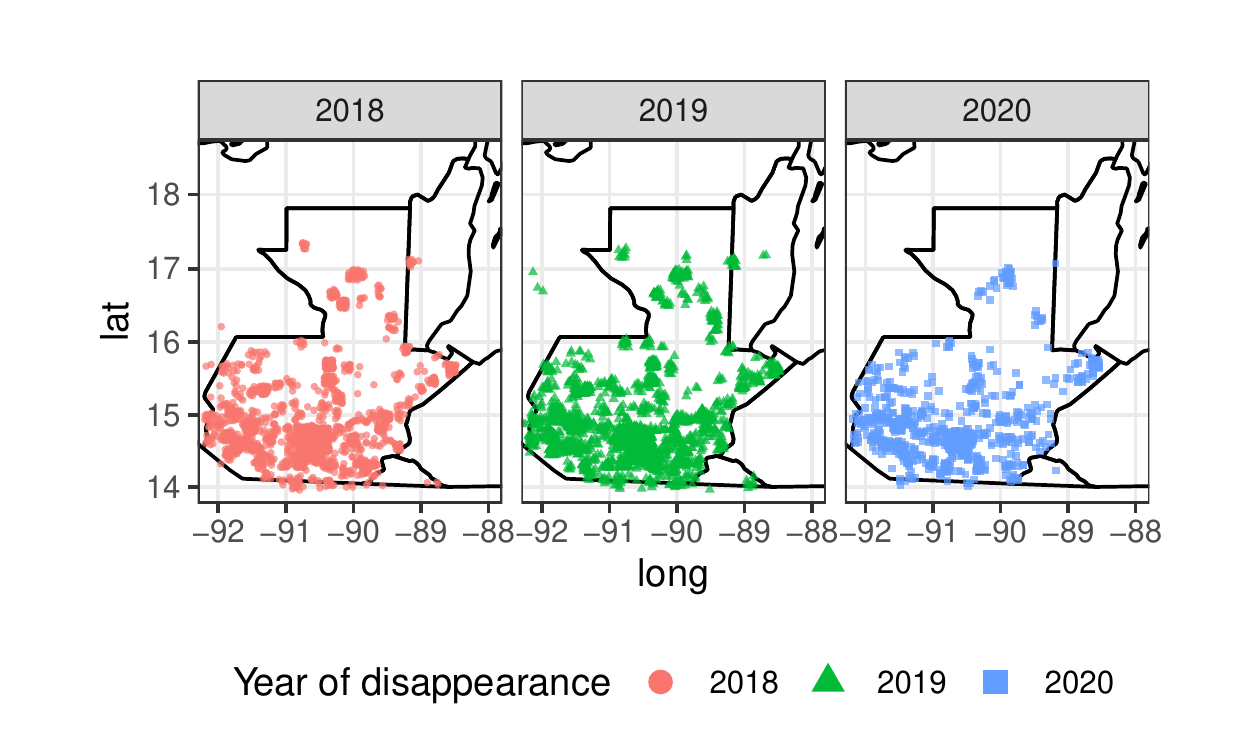}
    \caption{Geographic distribution of the number of reported missing children (2018 - 2020) according to the \textit{Alerta Alba-Keneth} Twitter data. Location data were extracted from the ``Place of disappearance'' field of the tweet.}
    \label{fig:twitter-map}
\end{figure*}

Next, we consider the geographic distribution of the disappearance events.
We obtained this crucial piece of information from the Twitter data since equivalent individual-level information is not shared by the National Civilian Police. 
Figure~\ref{fig:twitter-map} shows the location of the disappearance of missing children reported by the \textit{Alerta Alba-Keneth} Twitter account between 2018 and 2020. 
On the map, each individual disappearance is represented as a dot. 
Note that dots may not correspond to the exact locations since we applied a small degree of random noise to the location data (less than 0.001\degree) to avoid overplotting in areas with a high concentration of missing reports.

We highlight four important spatial patterns from this figure. 
First, missing children reports are concentrated in urban centers, more prominently in the capital city, which is home to about 20\% of the country's population. 
Second, we find a smaller concentration of missing reports in the country's Highlands, located in the Central and Western parts of the country. 
This is the region where most of the indigenous population, which correspondents to approximately half of the country's total, is concentrated.
It is also the region with the highest rates of poverty in the country, according to data from the National Institute of Statistics\footnote{\href{https://www.ine.gob.gt/ine/vitales/}{https://www.ine.gob.gt/ine/vitales/}}. 
Third, we find a higher concentration of missing cases in the south of the country. 
This area is mostly populated by non-indigenous groups and is generally characterized by higher levels of violence.
Fourth, we document a high incidence of disappearance events close to the country's boundaries. 
This is true for the eastern border with Mexico, the northwestern border with Belize, and the southeastern border with Honduras and El Salvador.
In addition to this, we find a high number of disappearances in the country's only port on the Atlantic Ocean (Puerto Barrios, in the west of the country). 
These are all regions characterized by high levels of insecurity and in which bands of human traffickers are known to operate. 
These are worrying patterns that deserve more attention. 

We do not have access to the number of children that are found after having been reported as missing from the \textit{Alerta Alba-Keneth}. 
Nevertheless, there are reasons to believe that the number of children that are found after having gone missing is very small.
According to police data obtained from this study, the number of children that are found constitutes around 5\% of the cases reported every year, on average.   
In addition, parents may not notify the authorities of a child's ``re-appearance'' for a number of reasons, including fear of losing custody of the child (Evelyn Espinoza, personal communication 19 September 2019).

\section{Discussion}
\label{sec:discussion}

The population of missing children is a relevant social issue in many societies, particularly in those where the levels of violence, drug, and human trafficking are high. Due to the lack of detailed information on this population, our study leveraged Twitter data in order to characterize missing children by age, sex, and place of disappearance. Here, we presented results from Guatemala, however, the methodology for collecting data from tweets can be replicated for other countries.

Our results showed a remarkable difference in the number of disappearances by sex and age. Female adolescents are the most likely group to go missing. The number of missing girls was about two times higher than that of boys, and around 75\% of missing girls were between 13 and 17 years old. 
We provide the first set of estimates on the specific sex and age of individuals more likely to be reported as missing in Guatemala by leveraging real-time and individual-level data from a social media platform.
By highlighting the striking number of missing girls aged 13-17, our results suggest that sex trafficking is a potential mechanism behind disappearances in Guatemala, which corroborates with the known link between disappearances and human trafficking \citep{rodas_andrade_informe_2021}. In Guatemala, 45\% of the detected victims of trafficking in 2019 are girls. Of the female victims, 70\% are between 14 and 17 years old \cite{united2021global}. 

However, due to the high levels of violence, particularly among men which, according to the National Police in Guatemala, correspond to 86\% of the victims of homicide in 2020\footnote{\url{https://infosegura.org/seccion/guatemala/}}, one might see our results by sex as surprising. If violence was the main mechanism behind missing children, boys would have gone missing more than girls. Our results showed the opposite. A possible explanation for this might be the age range we analyzed here. Violent deaths are less concentrated among children (10-17 years old) when compared to individuals aged 18-30\footnote{\url{https://infosegura.org/2022/02/04/homicidios-guatemala-2021/}}. Thus, our results suggest that violence might not be the main mechanism behind disappearances in Guatemala.

This study also showed the spatial patterns of disappearance. Our findings revealed a higher number of reported cases in urban centers. The regions with higher levels of poverty and where most are indigenous had lower numbers of missing children. This result may reflect a potential limitation of our study: the level of reporting cases may vary across the country.
A lower presence of State institutions in the poorest areas may reduce the number of reported missing cases.

Our study has a  number of limitations. 
First, the data we present, both from Twitter and from official sources, are reliant on missing reports. 
We are unable to observe cases of missing children that were not reported.
Second, we are not able to establish a comparison between the datasets on an individual-level, once only the Twitter data are available on an individual-level. In this case, we cannot identify the subset of disappeared children that appear in both datasets, or those cases that were not reported to the police, for example, but appear in the Twitter data.
Third, it is possible that the \textit{Alerta Alba-Keneth} Twitter account may not have tweeted all the missing person cases reported to them. 
Fourth, our automated geocoding is subject to a small degree of error arising from ambiguous place names (e.g., incorrect spelling). 
In future work, we plan to use regular expressions to identify other cases of disappearances discussed on Twitter, apart from those reported to the authorities or tweeted by \textit{Alerta Alba-Keneth}.

\section{Conclusion}
\label{sec:conclusion}

In this paper, we showed how digital data can be of great help to better understand social problems. Our study provided a detailed picture of missing children in Guatemala. We showed that the female adolescents are the most likely group to go missing, suggesting that sex trafficking is a potential mechanism behind disappearances in Guatemala given the high concentration of human traffic among girls between 14 and 17 years old. We propose that, in order to reduce the incidence of disappearances in Guatemala, authorities must improve protection for young girls. Moreover, greater efforts are needed to collect data on missing children in low-income settings, where missing children are a particularly prevalent and detrimental social problem.

\bibliography{refs}
\bibliographystyle{ACM-Reference-Format}

\end{document}